# Time-Resolved Measurements of Cumulative Effects in Gas Dynamics Induced by High-Repetition-Rate Femtosecond Laser Filamentation

ROBIN LÖSCHER[1, †], MALTE C. SCHROEDER[1, †*], ALAN OMAR[1],
AND CLARA J. SARACENO[1, 2]

[1]*Photonics and Ultrafast Laser Science, Ruhr-Universität Bochum, Universitätsstr. 150, 44801 Bochum, Germany*
[2]*Research Center Chemical Science and Sustainability, University Alliance Ruhr, Universitätsstr. 150, 44801 Bochum, Germany*

*†These authors contributed equally to this work.*
*\*malte.schroeder-w9x@ruhr-uni-bochum.de*

**Abstract:** The advent of high-average-power, ultrafast ytterbium-based lasers allows us to generate laser filaments at repetition rates ranging from 10s of kHz up to 100s of kHz. At such high repetition rates, the inter-pulse time lies below the time required for the total diffusion of the deposited heat by each laser pulse, leading to cumulative hydrodynamic effects that have so far been rarely studied. Here, we present, to the best of our knowledge, the first experimental time-resolved measurements of these dynamics in air for laser repetition rates between 1 kHz and 100 kHz. We measure the change in the air refractive index caused by the localized heat deposition and the length of the filament-generated plasma channel, with which we can infer the corresponding change in air density. We observe that at repetition rates above 10 kHz, stationary density depletions with vanishing dynamics emerge. Our findings are of wide relevance for the fields of high-repetition-rate laser filamentation and its applications, as well as THz generation from laser-induced plasma sources.

Femtosecond laser filamentation in gases and the effect of the associated self-induced long-timescale gas dynamics have been studied extensively in the past decades [1,2]. The formation of laser filaments requires peak powers on the scale of $10^9$-$10^{10}$ W, sufficient to overcome diffraction by self-focusing, which leads to the ionization of the medium and, thus, plasma formation. In the wake of filament-induced plasma generation, a transient extended channel of heated gas remains. The importance of this heated channel has been demonstrated for applications such as guiding and triggering of electrical discharges [3,4], lightning triggering [5], optical communication through fog [6,7], and waveguiding of continuous-wave laser beams over distances up to 50 meters [8,9]. The gas dynamics related to this localized heating have been investigated and well-documented for previous generations of ultrafast laser systems, where laser repetition rates from the Hz range and up to 1 kHz dominated the field. Here, the inter-pulse timing allows for an almost full recovery of the gas medium via thermal diffusion before the arrival of the subsequent pulse. Thus, the long-timescale hydrodynamics caused by filament plasmas could be studied and treated as an isolated single-shot effect [2,10,11], and their transient nature could not be exploited easily. With recent advances in ytterbium-based high-power ultrafast laser systems, a new regime, where the repetition rates of high energy laser systems capable of filamentation reach 10s to 100s of kHz, has become accessible. In this regime, the resultant shorter inter-pulse time does not allow for the complete diffusion of the residual heat left by the plasma recombination process. This results in cumulative effects in the gas hydrodynamics, which are critical to study, as they impact experiments employing high-repetition-rate, high-power, ultrafast laser systems as a driver for filamentation in gases.

In an early study by Koulouklidis *et al.* on terahertz generation in a filament plasma, the negative impact of the inter-pulse gas dynamics on the conversion efficiency at repetition rates up to 6 kHz has been shown [12]. In 2021, Walch *et al.* demonstrated cumulative effects in the air-density depletion induced by femtosecond laser filamentation and the potential for enhanced electrical discharge triggering [13]. Simulations of the hydrodynamics of the heated gas channel predicted a modulated, stationary gas depletion along the channel for repetition rates ≥ 20 kHz [14], which was attributed to the slow thermal diffusion of the air causing the formation of under-dense channels with lifetimes of up to about 90 ms [15]. In 2023, a stationary density depletion has been observed experimentally at repetition rates 40 kHz ≥ $f_{\text{rep}}$ ≥ 100 kHz, but without a time-resolved measurement of the hydrodynamics, their impact on spark gap triggering remained unclear at the time [16]. Moreover, recent studies demonstrated the suitability of filament plasmas for THz generation at repetition rates ≥ 6 kHz [17,18], with a potential correlation between the conversion efficiency and the induced gas hydrodynamics [19].

In this study, we investigate the time-resolved hydrodynamics of femtosecond laser filaments at repetition rates ranging from 1 kHz to 100 kHz by measuring the localized, heat-induced change in the air refractive index in the wake of the filamentation process via interferometry. The experimental setup is shown in Figure 1. We use a commercial regenerative amplifier (Light Conversion, *CARBIDE*) with pulse energies of 400 µJ and a maximum output power of 40 W at a central wavelength of 1.03 µm. A pulse picker allows the selection of different repetition rates from 100 kHz to 1 kHz. The pulses are split with 300 µJ pulses being used for the filamentation experiment. These initial pulses with a duration of 220 fs (FWHM) are compressed down to 50 fs in a home-built multi-pass cell (MPC) [20], reducing the pulse energy to 273 µJ, and reaching a peak power of 3.8 GW at all available repetition rates. Filamentation is accelerated by focusing the pulses with a lens ($f_p$ = 100 mm), resulting in the formation of a plasma channel. The plasma channel length was measured at different repetition rates via a capacitive plasma probe (CPP) [21]. Here, a pair of electrodes, one on either side of the plasma channel, is translated along the beam propagation axis, and the ionic charge is measured as a function of the electrode position. The measured ionic current corresponds to the spatially resolved relative plasma density [21]. We estimate the plasma channel length $L$ by extracting the length at the $1/e^2$ level. An example measurement is shown in the inset of Figure 1. From these measurements, we could determine the lengths of the created plasma channels to be, on average, about 19.3 mm, dependent on the laser repetition rate, with a variance of approximately 1.2 mm based on the chosen laser repetition rate. The CPP measurements provide us greater precision when determining the plasma channel length when compared to photographic length measurements, where the recorded fluorescent signal is weak at low plasma densities, resulting in a greater uncertainty due to insufficient signal-to-noise ratios. Higher repetition rates led to a reduction in plasma channel length due to a reduced Kerr nonlinearity caused by the steady-state density depletion but still yielded a plasma channel length that exceeded the estimated Rayleigh length of the focused laser beam of 7.7 mm. Special care had to be taken for the measurement at the 1 kHz repetition rate, where the CPP method was insufficient due to the low ionic current. For this case, the average length was assumed together with an uncertainty to account for possible variance in the length based on the laser repetition rate.

A pump-probe setup is used to measure the change in air density due to the heat deposited in the air after the plasma recombination. As a probe, we employ a digital-delay-triggered diode laser at a central wavelength of 450 nm and emitting nJ-pulses of 6 ns duration. As the heating changes the air refractive index locally, the wavefront of the collimated probe beam accumulates a spatially dependent phase shift by counter-propagating through the filamentation zone. We acquire fringe-encoded images of the phase shift by relay-imaging the probe beam through a folded-wavefront interferometer (FWI) onto a camera [2,16,22]. By changing the delay between the filamentation and probe pulses, time-resolved measurements of the air density evolution are performed [2].

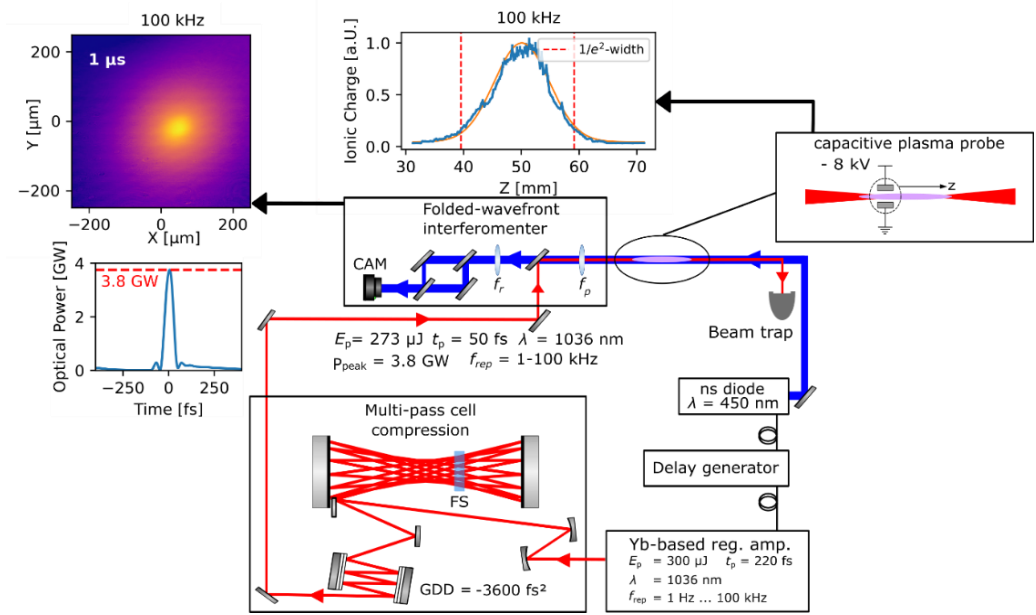

**Figure 1.:** Schematic representation of the experimental setup. The compressed pulses from the MPC are focused at $f_p$ = 100 mm to generate the plasma channel. The focal plane of the $f_p$-lens is relay imaged using $f_r$ = 500 mm through the folded-wavefront interferometer (FWI) onto a camera. The FWI is used for the phase retrieval, and the CPP setup is used to measure the filament length. The first inset shows a typical phase map retrieved at 1-µs delay after plasma generation via the FWI at a laser repetition rate of 100 kHz. The second inset shows a CPP measurement, where the measured ionic charge is plotted over the filament propagation distance. Marked in red is the $1/e^2$ signal decay, which determines the length of the filament plasma [16]. The acronym FS stands for fused silica and GDD for group velocity dispersion.

It is worth noting that this method depends on excellent shot-to-shot stability regarding the interferometric image and visibility of the fringes on the camera. High stability allows for a rolling shutter camera, which carries the benefit of larger chip sizes and, thus, high spatial resolution. In the presented experiment, we used the ZWO *ASI 6200/M* camera, which employs the SONY *IMX455* chip with a resolution of 9576x6388 pixels and a pixel size of 3.76 μm, corresponding to (166±3) nm calibrated pixel size based on our imaging setup. This is made possible by the high stability of the pump and probe lasers and the low camera read noise. The acquired images can be analyzed by a common Fourier-transform algorithm to retrieve the accumulated phase of our probe [23].

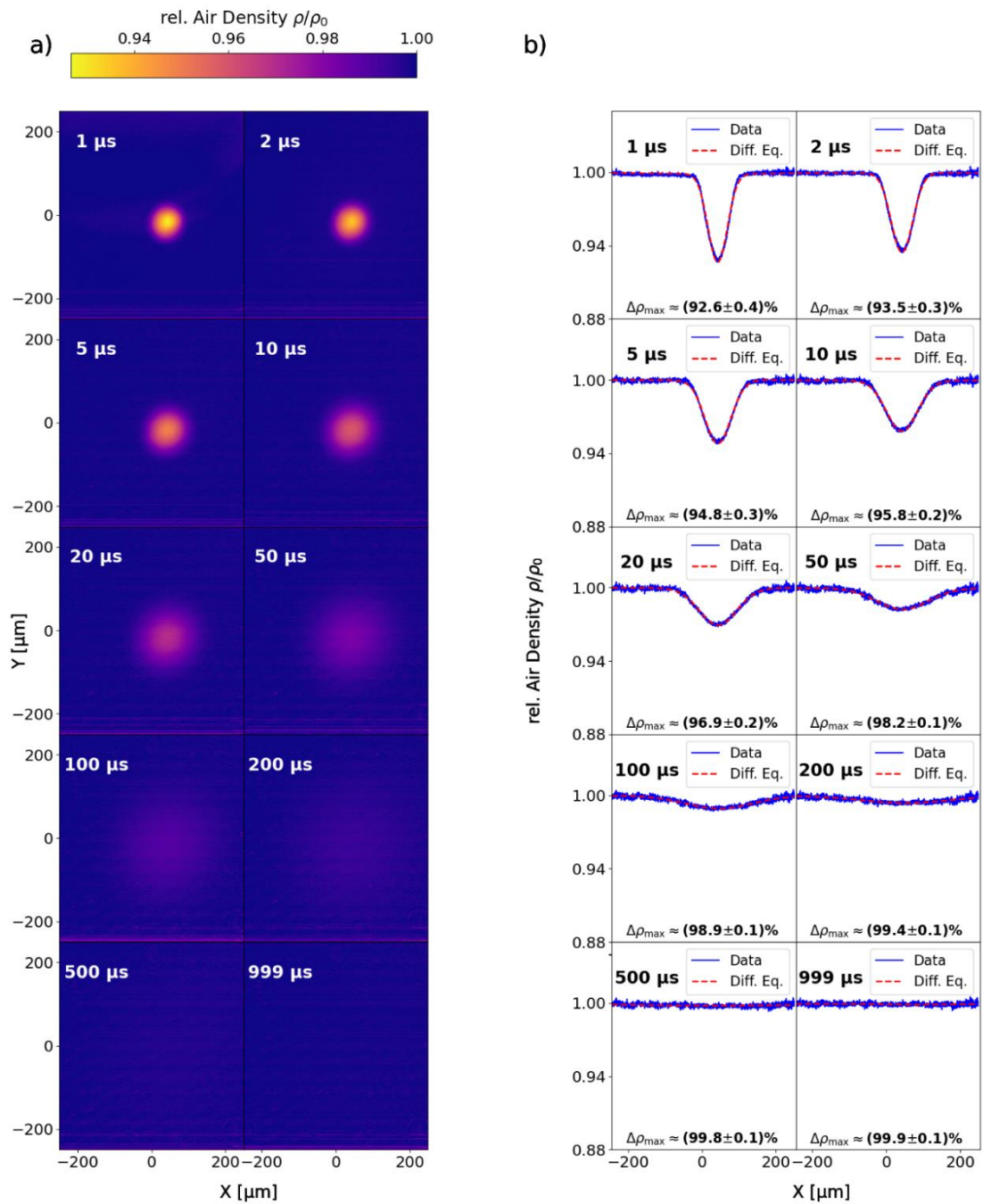

**Figure 2.:** 1 kHz case. Shown in a) is the recovered density depletion in air from the image plane in the filamentation region at different time delays after the incident pump. In b), lineouts of the depletion profiles along the maximum depression are shown in blue to visualize the evolution of the density depletion due to diffusion and convection. The data was fitted following a thermal diffusion equation (see Eq. 1), shown in red, to retrieve the maximum and minimum depletion [2].

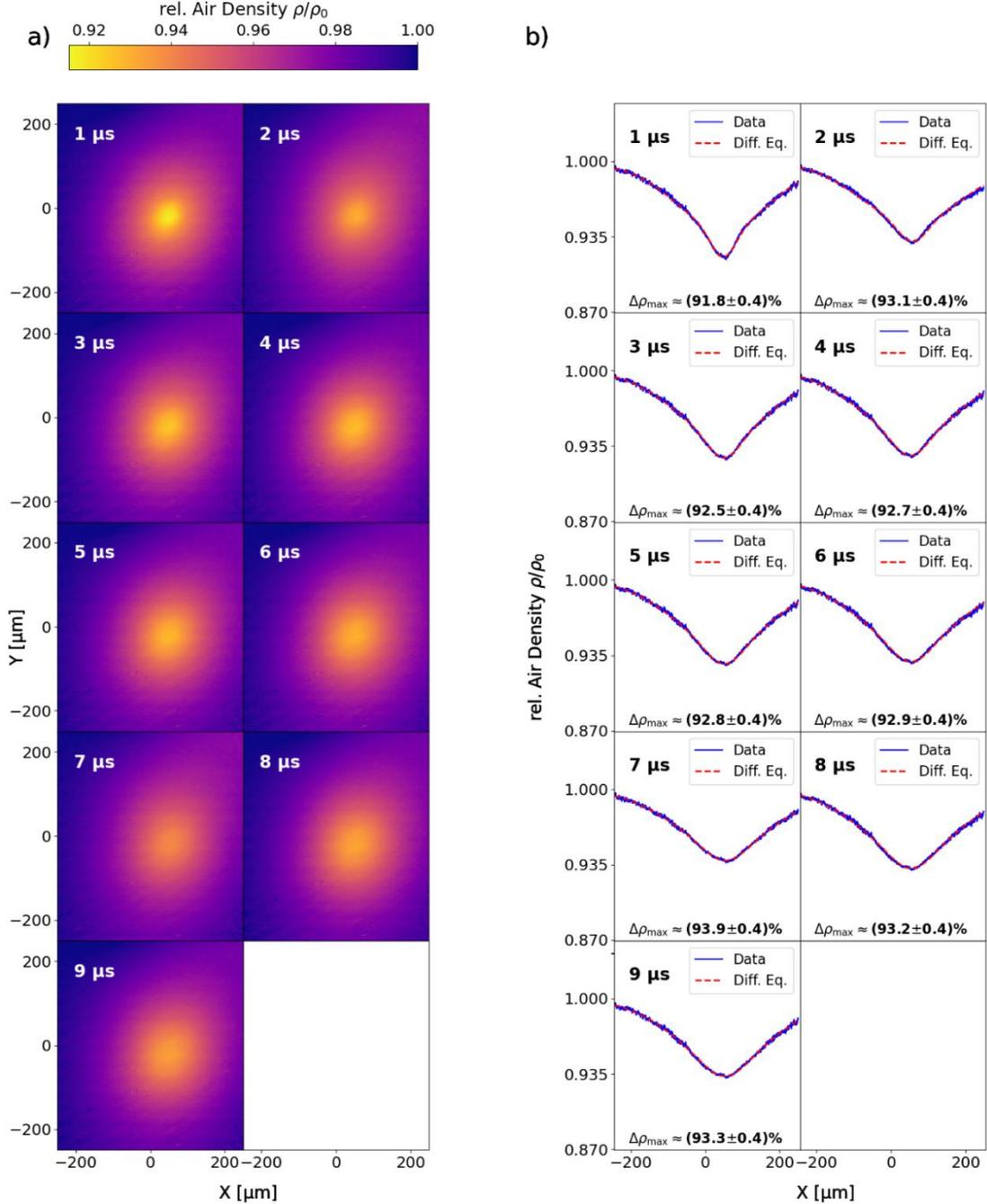

**Figure 3.:** 100 kHz case. Shown in a) is the recovered density depletion in air from the image plane in the filamentation region at different time delays after the incident pump. In b), lineouts of the depletion profiles along the maximum depression are shown to visualize the evolution of the density depletion due to diffusion and convection. The data was fitted following a modified thermal diffusion equation, shown in red, to retrieve the maximum and minimum depletion.

The algorithm allows the retrieval of the localized phase shift $\Delta\Phi(x,y,t)$ of the probe beam wavefront from the interferograms captured using the folded wavefront interferometer. A retrieved exemplary phase map is shown as an inset in Figure 1. As the phase shift $\Delta\Phi(x,y,t)$ is caused by the heating of the air in the plasma channel volume during the recombination process of the plasma, and the air refractive index scales inversely with the air temperature $n_{\text{air}} \propto 1/T_{\text{air}}$, a change in local air temperature will introduce a refractive index transient $\Delta n(x,y,t) = (k_0 L)^{-1}\Delta\Phi(x,y,t)$. Here, $k_0$ is the vacuum wavenumber of the probe pulse at the peak of its spectrum and $L$ is the $1/e^2$-length of the filament plasma channel. The transient $\Delta n(x,y,t)$ perturbs the wavefront of the ns-probe beam when it propagates through the air containing the heated air volume, resulting in the measurable phase shift $\Delta\Phi(x,y,t)$. With the lengths of the plasma channel we measured via the CPP method, information about the changes in the local air refractive index can be gained from these phase shifts. In turn, as the air refractive index follows $n_{\text{air}} \propto \rho_{\text{air}}$ in relation to the air density, we calculate the relative depletion in the local gas density $\rho(x,y,t)$ from the change in the refractive index $\Delta n(x,y,t)$.

Interferograms were recorded for laser repetition rates of 1 kHz, 10 kHz, 20 kHz, 33 kHz, 50 kHz, and 100 kHz. Ten interferograms were taken for each repetition rate, and several minutes between measurements were taken to show the repeatability of the method and allow for statistical evaluation. As examples, Figures 2a) and 3a) show the density change $\Delta\rho(x,y,t)$ caused by the plasma heating of the air at different time delays after the plasma generation for laser repetition rates of 1 kHz and 100 kHz, respectively. The temporal evolution of the heat-induced density hole was probed between two pump laser pulses at different time delays ranging from 1 µs to 999 µs after the incident pulse for the 1 kHz case and between 1 µs to 9 µs for the 100 kHz case. Figures 2b) and 3b) show corresponding lineouts of the gas density shown in Figures 2a) and 3a). The lineouts are taken across the maximum density depletion at each time delay. With the assumption that the initial temperature distribution along the heated volume is Gaussian in profile, the depletion data is fitted with a solution to the thermal diffusion equation,

$$T(x,y,t) = T_0 \left( \frac{R_0^2}{R_0^2 + 4\alpha t} \right) \exp\left( -\frac{\sqrt{x^2+y^2}^2}{R_0^2 + 4\alpha t} \right) + T_{\text{b}}, \tag{1}$$

with $T_0$ being the maximum temperature, $\alpha$ the thermal diffusion coefficient of air, $R_0$ the initial heated channel radius, $t$ the time, and $T_{\text{b}}$ the background temperature, to retrieve the maximum density depletion for a laser repetition rate of 1 kHz. As previously shown, this equation yields accurate results when applied to gas heated through laser filament-induced plasma generation [2]. For higher laser repetition rates, the equation evolves into a series of thermal diffusion equations $\sum_n T(x,y,n\cdot t)$. This is required to account for heat diffusion and the cumulative nature of the recorded density depression, where deposited heat in the air, remaining from previous laser pulses, affects the subsequent ones. For time scales in the 10-µs to 1-ms range and beyond, we can assume that, apart from the heat deposition during the plasma recombination, heat diffusion dominates as the driving mechanism behind the observed hydrodynamic profiles [2]. The air density depletion profiles show some marked differences based on the repetition rate of the pump laser. Considering the two presented cases, 1 kHz and 100 kHz, we can observe differences in the gas density depletion profiles in the profile sequences shown due to the absence and presence of heat accumulation.

At 1 kHz, averaged over all 10 interferograms, the initial depletion profile at 1 µs after plasma generation reaches ~92.6 % with a width of about 62 µm (FWHM). Over time, the gas density hole widens and becomes shallower, resulting in a air density reduction to only ~95.8 % of after 10 µs, and a width of approximately 99 µm (FWHM). The depleted region is replenished from the surrounding ambient air over a few hundred microseconds, and only a residual density reduction to ~99.4 % at 200 µs remains. At 1 µs before the next laser pulse arrives, the air density is recovered to ~99.9 %. As such, the subsequent pulse interacts with ambient air showing a homogenous density profile across the interaction volume. Since the density holes are replenished before the next pump pulse arrives, no accumulations occur as not enough energy is deposited in the air medium quickly enough to generate an extended and permanent reduction in the local gas density. These strong modulations of the local refractive index is the result, which can affect secondary collinear probe or signal beams [7,24].

For the 100-kHz case, the initial profile at 1-µs delay shows a wide depletion, spanning approximately 193 µm (FWHM) and reaching ~91.8 % at the maximum. The maximum lies within a bump, a transient depression in the gas density caused by the most recent pump pulse, at the apex of a much broader gas depletion well. While the transient rapidly dissipates, reducing the maximum depletion to ~93.1 % after 2 µs, the depletion well retains its expansion of approximately 231 µm (FWHM) and does not fall below a depth of ~92.1 % in the time between two consecutive pump pulses. Therefore, a steady state with a continuous density depression is achieved, extending beyond the heated volume's boundaries at lower repetition rates, as shown in Figure 2.
For other repetition rates (see Supplementary Figure S2-5), the results appear quantitively similar. The experiments were repeated for the repetition rates of 10 kHz, 20 kHz, 33 kHz, and 50 kHz. The evolution of the maximum depletion is shown in Figure 4 for all investigated laser repetition rates. The initial maximum depletion reaches similar levels at all repetition rates of around 90 % with different amounts of recovery between shots due to the different intra-pulse times, but also different recovery rates over the same timeframes. This is most likely due to the expansion of the depleted domain, or formation of a "well", at higher repetition rates. The density hole is filled from the surrounding air volume, with the flow rate determined by the pressure gradient between the depleted and the ambient air volume. At low repetition rates, the heating of the air is abrupt, leading to a sharper and deeper density depression. At higher repetition rates, the formation of the well leads to a more gradual profile, which slows down the recovery speed. This behavior aids in the formation of continuous density depletion. It appears that the accumulation effects become more pronounced with increasing repetition rates, clearly visible at a repetition rate of 10 kHz, but are already significant at repetition rates exceeding 1 kHz [12].

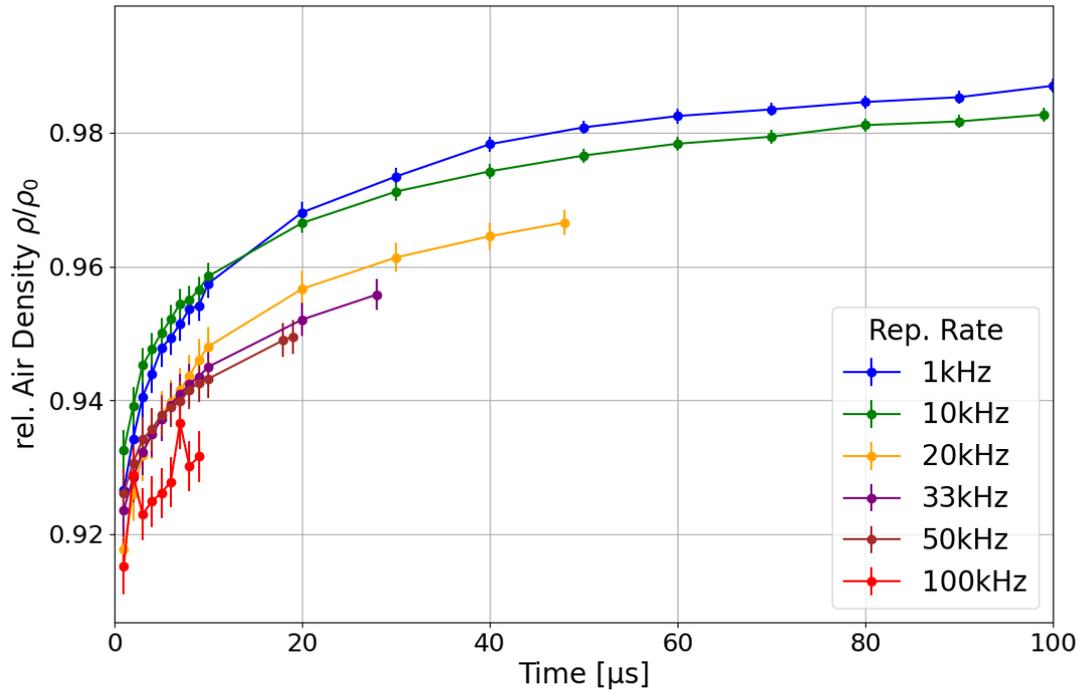

**Figure 4.:** The figure shows the evolution of the maximum density depletion over time for the different repetition rates of the pump laser system. Only the first 200 µs are shown to aid visualization. The figure showing the evolution over the maximum timescale of 1 ms can be found in the supplementary (Figure S1).

In summary, the results shown in this letter indicate that at higher laser repetition rates, laser filamentation can be used to remotely reduce local gas density in a permanent fashion. While the initial depletion after the plasma recombination shows little variance, the formation of a permanent depletion well with decreasing modulations at higher repetition rates opens up interesting applications, such as the improved capacity for the guiding of electrical discharges [14,16] as well as the generation of permanent air waveguides that are not constrained by the diffusion time of the deposited heat in the air [8,25]. Conversely, due to the reduction in the pressure gradient created by the heat deposition during plasma recombination, the amplitude of the resultant pressure wave is reduced in turn. This might result in diminished efficacy when considering high-repetition-rate filamentation for applications such as cloud clearing for free-space optical telecommunication [6,26]. The effect of a permanent gas density reduction of the propagation medium would also affect the process of laser filamentation itself [27,28]. As the non-linear refractive index is dependent on the temperature and ambient pressure of the propagation medium, a permanent, localized temperature increase and consequent pressure reduction along the propagation axis of the ultrafast laser pulses will diminish the occurring nonlinearities, potentially leading to a delay in the onset of filamentation. At this point, further study is required to investigate how this effect scales with higher pulse energies and whether this effect could be used to coherently extend the filamentation length by modulating the laser repetition rate. The next steps would also lead to investigations at repetition rates exceeding 100 kHz (below 10-µs inter-pulse timing), as new high-power, MHz laser oscillators [29] are emerging, which in combination with secondary compression stages, like multi-pass cells [20], can provide ultrafast laser pulses, which, while insufficient for laser filamentation, are sufficient for plasma generation at MHz frequencies. Here we would rapidly approach a hydrodynamic regime that is no longer purely dominated by thermal diffusion but would force us to consider faster dynamics, such as the initial pressure wave generation caused by the rapid heat deposition during the plasma recombination step [2], as well as more complex plasma excitation and recombination pathways in the generated plasma itself [30].

**Funding**

Funded by the Deutsche Forschungsgemeinschaft (DFG, German Research Foundation) under Germanys Excellence Strategy – EXC-2033 – Projektnummer 390677874 - RESOLV. The research was conducted in the Research Center Chemical Sciences and Sustainability of the University Alliance Ruhr at Ruhr-Universität Bochum. We acknowledge support by the DFG Open Access Publication Funds of the Ruhr-Universität Bochum.

**Acknowledgments**

We thank Prof. Dr. Martina Havenith-Newen and Dr. Claudius Hoberg (Department of Physical Chemistry II, Ruhr-Universität Bochum) for their support.

**Disclosures**

The authors declare no conflicts of interest.

**Data availability**

Data underlying the results presented in this paper is publicly available on the Zenodo platform DOI: 10.5281/zenodo.14528906 [31]

**Supplemental document.** See Supplementary for supporting content.